\documentclass[english]{iopart}
\usepackage[T1]{fontenc}
\usepackage[latin9]{inputenc}
\usepackage{amssymb}
\usepackage{graphicx}
\usepackage{esint}

\makeatletter

\usepackage{iopams}
\usepackage{setstack}
\renewcommand\footnoterule{%
  \kern-3\p@
  \hrule\@width.4\columnwidth
  \kern2.6\p@}
\renewcommand\@makefntext[1]{%
    \parindent 1em\noindent
    \hb@xt@1.8em{\hss$^{\@thefnmark}$)}\hspace{2pt}%
    \footnotesize\rmfamily#1}  
\def\@makefnmark{\hspace{.5pt}\hbox{$^{\@thefnmark}$%
\hspace{-1pt}}}
\makeatother

\usepackage{babel}
\begin{document}

\title{Time-dependent Hamiltonians with 100\% evolution speed efficiency}

\title[Time-dependent Hamiltonians with 100\% evolution speed efficiency]{}

\author{Raam Uzdin$^{1}$, Uwe Günther$^{2}$, Saar Rahav$^{3}$ and Nimrod
Moiseyev$^{1,3}$}

\address{$^{1}$Faculty of Physics, Technion - Israel Institute of Technology\\
$^{2}$Helmholtz-Center Dresden-Rossendorf, POB 510119, D-01314 Dresden,
Germany\\
$^{3}$Schulich Faculty of Chemistry, Technion - Israel Institute
of Technology\\
\eads{\mailto{raam@tx.technion.ac.il},\quad{}\mailto{u.guenther@hzdr.de}} }
\begin{abstract}
The evolution speed in projective Hilbert space is considered for
Hermitian Hamiltonians and for non-Hermitian (NH) ones. Based on the
Hilbert-Schmidt norm and the spectral norm of a Hamiltonian, resource-related
upper bounds on the evolution speed are constructed. These bounds
are valid also for NH Hamiltonians and they are illustrated for an
optical NH Hamiltonian and for a non-Hermitian $\mathcal{PT}-$symmetric
matrix Hamiltonian. Furthermore, the concept of quantum speed efficiency
is introduced as measure of the system resources directly spent on
the motion in the projective Hilbert space. A recipe for the construction
of time-dependent Hamiltonians which ensure 100\% speed efficiency
is given. Generally these efficient Hamiltonians are NH but there
is a Hermitian efficient Hamiltonian as well. Finally, the extremal
case of a non-Hermitian non-diagonalizable Hamiltonian with vanishing
energy difference is shown to produce a 100\% efficient evolution
with minimal resources consumption.\global\long\def\ket#1{\left|#1\right\rangle }
 \global\long\def\bra#1{\left\langle #1\right|}
 \global\long\def\braket#1#2{\left\langle #1|#2\right\rangle }

 \global\long\def\ketbra#1#2{\left|#1\right\rangle \left\langle #2\right|}
 \global\long\def\braOket#1#2#3{\left\langle #1\left|#2\right|#3\right\rangle }
 \global\long\def\ra{\rangle}
 \global\long\def\la{\langle}
 \global\long\def\cH{\mathcal{H}}
 \global\long\def\cP{\mathcal{\mathcal{P}}}
 \global\long\def\cT{\mathcal{T}}
 \global\long\def\fH{\mathfrak{H}}
 \global\long\def\RR{\mathbb{R}}
 \global\long\def\CC{\mathbb{C}}
 \global\long\def\PP{\mathbb{P}}
 \global\long\def\dd{\dagger}
 \global\long\def\a{\alpha}
 \global\long\def\b{\beta}
 \global\long\def\sg{\sigma}
 \global\long\def\t{\theta}
 \global\long\def\T{\Theta}
 \global\long\def\lb{\lambda}

\end{abstract}

\section{Introduction}

Non-Hermitian models naturally emerge in many fields of physics as
efficient tools for the description of complicated large systems in
terms of smaller effective subsystems \cite{Nimbook,Berry NH phys}.
Examples range from atomic/molecular physics \cite{Gilary,Berry grating},
light propagation in optically active crystals \cite{MVB_pol} and
media with anisotropic pumping and absorption \cite{christo-prl-08,KGM-prl-08,kottos-prl-09,christo-prl-09,Longhi-prl-09,christo-nature-10,Kottos-nature-10,PT-lattice-pra-10,eva-hugh-pra-11,Morales,WG}
over microwave cavities \cite{Richter EP,pt-micro-prl-2012}, coupled
electronic circuits \cite{kottos-LRC-pra-11,kottos-UG-LRC-pra-12}
up to mechanics \cite{Seyranian-Mailyb-book,cmb-mech-exp-2012}, hydrodynamics
\cite{schmid-book-2001,shkalikov-2004} and magnetohydrodynamics \cite{GSZ-jmp-2005,GK-jpa-2006,GLT-2010,stefani-spectra-2012}.
Apart from the spectral properties of the Hamiltonians, the evolution
processes generated by non-Hermitian (NH) Hamiltonians can significantly
differ from those generated by Hermitian Hamiltonians \cite{MVB_pol,cmb-brach-prl-2007,cmb-lnp-2009,BU,UMM}.
In this regard it appears natural to ask what new possibilities NH
evolution entails and what bounds can be broken when a Hamiltonian
is no longer Hermitian.

For example, it was shown in \cite{cmb-brach-prl-2007} that a NH
$2\times2$ matrix Hamiltonian with some predefined energy difference
can generate a much faster evolution than a Hermitian Hamiltonian
with the same energy difference. The evolution speed is the rate in
which a state changes into other states (e.g. the angular speed of
the state vector on a Bloch sphere). In the Hermitian case, the energy
difference sets an upper bound on the evolution speed (the Fleming
bound \cite{Fleming}). The implicitly demonstrated clear violation
of this bound by non-unitary evolution processes of NH systems leads
to the conclusion that these energy-difference based bounds should
be replaced by some more adequate bounds for NH systems.

The first goal of this article is to derive an upper bound on the
evolution speed that works for any Hamiltonian, be it Hermitian or
non-Hermitian. The bounds derived here may not be tight for some Hamiltonians
and/or for some initial conditions, but this statement is equally
true for the Hermitian Fleming bound.

This leads directly to the second goal of this article: to show how
to construct Hamiltonians for which the evolution speed of the state
of interest reaches the upper bound for any instant of the evolution.
We call these Hamiltonians {}``maximal efficiency\textquotedblright{}
Hamiltonians (or maximally efficient Hamiltonians). For every state
evolution there exists a family of Hamiltonians that are maximally
efficient. This family contains both Hermitian and NH Hamiltonians.

Our third goal is to explore a very special Hamiltonian in this family
which is of rank-one, non-diagonalizable and similar to a Jordan block
with zero-eigenvalue. This Hamiltonian corresponds to a NH degeneracy
called exceptional point (EP) which has only one (geometric) eigenvector\textbf{
}\cite{Nimbook,Seyranian-Mailyb-book,baumg}.\textbf{ }We show that
any state evolution can be generated solely by such NH degeneracies
yielding an EP-driven evolution. This special evolution minimizes
the Hilbert-Schmidt norm $\sum_{i,j}|\cH_{i,j}|^{2}$ of the matrix
Hamiltonian $\cH$.

We note that the second goal strongly differs conceptually from the
so called quantum brachistochrone problem. The quantum brachistochrone
problem consists in finding the (time independent) Hamiltonian which
evolves some predefined initial state into some predefined final state
in a minimal time. This problem was the subject of intensive studies
during the last few years for Hermitian systems \cite{Dorje Herm,Hook,japan-brach-prl-2006}
as well as for NH ones \cite{cmb-brach-prl-2007,cmb-lnp-2009,Assis,GRS-ep-jpa-2007,Ali-brach-prl-2007,Giri,GS-brach-pra-2008,GS-brach-prl-2008,Ali-brach-pra-2009}.
As shown in \cite{Dorje Herm,Hook,AA-prl-1990} for Hermitian systems,
the corresponding minimal-passage-time trajectories correspond to
geodesics in projective Hilbert space. For the evolution problems
we are investigating here, the trajectories in projective Hilbert
space are predefined and not necessarily geodesic. Instead, we are
searching for Hamiltonians capable of producing evolution processes
which exactly follow these predefined trajectories with minimal resources.
That is, we look for efficient evolution and not for a fast evolution.
In fact, our optimization problem is closer in spirit to the reverse
engineering approach used to quicken adiabatic evolution \cite{Demir,MVB Trnsls,Muga}.
Yet there are two main differences: The first difference is that in
our case we seek only Hamiltonians which yield maximal efficiency.
The second difference is that we take as input only the evolution
of a single state (the state of interest), while in \cite{Demir,MVB Trnsls,Muga}
the number of states needed to be specified is equal to the Hilbert
space dimension (number of levels in the system).

The article is organized as follows: Section \ref{sec: Preliminaries}
contains some basic facts on the evolution speed in projective Hilbert
space. In section \ref{sec: bounds}, the Hilbert-Schmidt norm and
the spectral norm of a Hamiltonian are introduced as upper bounds
on the evolution speed. In section \ref{sec :Speed-Efficiency}, the
concept of speed efficiency is introduced, and for the predefined
evolution of a given state a family of Hamiltonians is constructed
which ensure a speed efficiency of 100\%. The generic properties of
maximally efficient evolutions are explored. Section \ref{sec: EP-DE}
is devoted to the special case of a maximally efficient evolution
which is driven by a Hamiltonian at an EP (a NH degeneracy). In Appendix
1, for completeness we briefly discuss the relation between Bloch
sphere and projective Hilbert space. In Appendix 2, the norm speed
bounds are illustrated for a Hamiltonian that describes two optical
systems recently studied. In Appendix 3, the norm speed bound is applied
to the matrix Hamiltonian of a $\mathcal{PT}-$symmetric quantum brachistochrone.

\section{Preliminaries - the evolution speed in projective Hilbert space $\PP(\fH)$\label{sec: Preliminaries}}

Let $\ket{\psi}\in\fH=\CC^{N}$ be a solution of the time-dependent
Schrödinger equation (TDSE):
\begin{equation}
i\partial_{t}\ket{\psi}=H(t)\ket{\psi},\label{eq: TDSE}
\end{equation}
where $H(t)\neq H^{\dd}(t)\in\CC^{N\times N}$ is the matrix of the
corresponding time-dependent NH Hamiltonian. Defining the bra-vector
$\la\psi|$ in a standard way%
\footnote{Unlike other choices often made for NH Hamiltonians in order to exploit
the bi-orthogonal relations of the eigenstates\textbf{ }\cite{Nimbook}.%
} as $\la\psi|=|\psi\ra^{\dd}$, the adjoint TDSE has the form
\begin{equation}
-i\partial_{t}\la\psi|=\la\psi|H^{\dd}(t)\,.\label{eq: TDSE-2}
\end{equation}

Our main interest is to study the rate at which states evolve into
different states. Phase evolution is irrelevant for this purpose.
It makes sense, then, to study the evolution of states in a space
where the phase is eliminated. The so called projective Hilbert space
(PHS) is exactly suited for this purpose. The well known Bloch sphere
for two-level systems is closely related to PHS (see Appendix 1),
but strictly speaking it is not a PHS. For the reader not familiar
with PHS we provide a simplified and very limited presentation of
the basic ideas needed to understand the present work. For a more
complete and rigorous treatment see, e.g., \cite{quant-geom-book}.

The angle $\Theta$ between two complex vectors $\ket{\psi_{1}},\ket{\psi_{2}}\in\mathbb{\mathbb{C}}^{N}$
can be obtained from the standard inner product of the two vectors:
\begin{equation}
\cos\Theta=\frac{\left|\braket{\psi_{1}}{\psi_{2}}\right|}{\sqrt{\braket{\psi_{1}}{\psi_{1}}}\sqrt{\braket{\psi_{2}}{\psi_{2}}}}\label{eq: angle}
\end{equation}
The angle $\Theta$ acts as a measure of distance between two states:
$\Theta=0$ means the two states are identical up to a complex factor
and $\Theta=\pi/2$ indicates the states are mutually orthogonal%
\footnote{This is different from the Bloch sphere construction discussed in
Appendix 1, where orthogonal states correspond to the angle of $\pi$
between antipodal points on the sphere.%
}.

Now imagine that a state is infinitesimally changed from $\ket{\psi}$
to $\ket{\psi}+\ket{d\psi}$ where $\braket{\psi}{\psi}\gg\braket{d\psi}{d\psi}$.
The angle between the original state and the modified state, $d\Theta$,
can be obtained from (\ref{eq: angle}). Keeping leading orders in
$\ket{d\psi}$ and $d\Theta$ we get:
\begin{equation}
d\Theta^{2}=\frac{\braket{d\psi}{d\psi}}{\braket{\psi}{\psi}}-\frac{\braket{d\psi}{\psi}\braket{\psi}{d\psi}}{\braket{\psi}{\psi}^{2}}=ds_{FS}^{2}.\label{eq: FS-metric}
\end{equation}

$ds_{FS}^{2}$ is known as the Fubini-Study metric \cite{quant-geom-book}
which describes the length of an infinitesimal arc traced on a unit
hypersphere by changing a state by $\ket{d\psi}$. To quantify the
rate at which states change we will look at the evolution speed defined
by: $\left|\frac{d\Theta}{dt}\right|$ (or equivalently $\left|\frac{ds_{FS}}{dt}\right|$),
which can be interpreted as angular speed/frequency. This hypersphere
is related to the projective Hilbert space associated with the Fubini-Study
metric. All states which differ by a complex number are mapped to
the same point on the hypersphere (hence phase is immaterial in this
space). The details of this mapping are not important for the present
article. What is important is that the distance between two states
on the hypersphere which differ by $\ket{d\psi}$ is given by (\ref{eq: FS-metric}).
Formally, the projective Hilbert space of an N-level system is denoted
by $\PP(\fH)=\CC\PP^{N-1}=\CC_{*}^{N}/\CC_{*}$ where $\CC_{*}^{N}=\CC^{N}-\{(0,0,\ldots,0)\}$,
$\CC_{*}:=\CC-\{0\}$. As the state $\ket{\psi}$ evolves in time
it traces a certain trajectory in $\PP(\fH)$ (i.e. on the hypersphere
associated with the PHS). We denote the trajectory induced by $\ket{\psi}$
by $\pi(\ket{\psi})\in\PP(\fH)$. Notice that for any complex function
of time, $c(t)$ $(c(t)\neq0)$:
\begin{equation}
\pi(\ket{\psi})=\pi(c(t)\ket{\psi})\label{eq: pi traj}
\end{equation}

Next we wish to establish a relation between the evolution speed $\left|\frac{ds_{FS}}{dt}\right|$
and the Hamiltonian. Making use of the TDSE (\ref{eq: TDSE}) and
its adjoint (\ref{eq: TDSE-2}), and introducing the normalized state
vectors $|\Psi\ra:=|\psi\ra/\sqrt{\la\psi|\psi\ra}$, we find the
squared evolution speed in $\PP(\fH)$ is given by:
\begin{equation}
\fl\left(\frac{ds_{FS}}{dt}\right)^{2}=\la\Psi|H^{\dd}(t)H(t)|\Psi\ra-\la\Psi|H^{\dd}(t)|\Psi\ra\la\Psi|H(t)|\Psi\ra=:K(t).\label{eq: FS K def}
\end{equation}
Henceforth, we refer to $K(t)$ as \textquotedbl{}kinetic scalar\textquotedbl{}
because it plays a structurally similar role like the kinetic energy
in classical mechanical systems. The expression (\ref{eq: FS K def})
is a straight-forward generalization for non-Hermitian Hamiltonians
of the corresponding evolution speed discussed in \cite{AA-prl-1990,quant-geom-book}
for Hermitian systems. We note that for those systems $K(t)$ just
reduces to the instantaneous energy variance $K(t)=\la\Psi|H^{2}(t)|\Psi\ra-\la\Psi|H(t)|\Psi\ra^{2}=\Delta E^{2}(t)$.\textbf{
}Further insight into this expression can be obtained by\textbf{ }introducing
an instantaneous orthonormal basis set $\{|k\ra\}_{k=1}^{N}\in\fH=\CC^{N}$,
with $|\Psi\ra$ identified with one of its elements $|\Psi\ra=|j\ra$,
the kinetic scalar (\ref{eq: FS K def}) in this basis set takes the
form:
\begin{equation}
K=\sum_{k=1}^{N}\la j|H^{\dd}|k\ra\la k|H|j\ra-\la j|H^{\dd}|j\ra\la j|H|j\ra=\sum_{k\neq j}|\la k|H|j\ra|^{2}\ge0.\label{eq:fub-3}
\end{equation}
Obviously, $K$ is characterizing the total rate for transitions from
the given state $|\Psi\ra=|j\ra$ to other states of the system. Splitting
off the trace of the Hamiltonian
\begin{equation}
H=\cH+\mu I,\qquad\mu:=\mbox{Tr}(H)/N\label{fub-4}
\end{equation}
one immediately sees that $K$ is invariant with regard to trace shifts%
\footnote{A time-dependent trace $\mbox{tr}(H)=N\mu(t)\in\CC$ can be removed
from the Hamiltonian $H$ by the transformation $\ket{\psi}\to e^{i\intop_{0}^{t}\mu(t')dt'}\ket{\psi}$.
Since this transformation involves only a multiplication by a complex
function the motion in the PHS is not affected by this transformation
(\ref{eq: pi traj}). Setups with non-vanishing complex traces have
been considered, e.g., in \cite{Assis}.%
}, $K[H]=K[\cH]$. Hence, we can subsequently restrict our attention
to traceless Hamiltonians $\cH$.\textbf{ }In Appendix 1, we discuss
the dynamics on the PHS-related Bloch sphere and obtain its relation
to the kinetic scalar.

\section{Upper norm bounds on the evolution speed\label{sec: bounds}}

The absolute values of the matrix elements of Hermitian or non-Hermitian
Hamiltonians are directly defined by the intensities of interactions
and the strength of the corresponding fields. Naturally, a model remains
valid only within the region of applicability of the corresponding
underlying theory and/or the applicability of the approximations made
in the derivation of the model.\textbf{ }Hence, it is natural to ask
for the maximal evolution speed achievable by a given quantum system
when the resources are limited.

\subsection{The Hilbert-Schmidt-norm upper bound on the evolution speed}

To obtain the first simple upper bound on the evolution speed we notice
that:
\begin{eqnarray}
K\le\la\Psi|\cH^{\dd}\cH|\Psi\ra\le\mbox{tr}(\cH^{\dd}\cH)\equiv||\cH||_{HS}^{2}.\label{eq: K<HS}
\end{eqnarray}
where $||\cH||_{HS}^{2}$ is the Hilbert-Schmidt norm of the Hamiltonian
\cite{horn}. It is also known as Euclidean norm, $l_{2}-$norm, Schatten
2-norm, Frobenius norm and Schur norm \cite{horn}. It was possible
to use the trace in the last inequality since $\mathcal{H}^{\dagger}\mathcal{H}$
is a positive operator. Equations (\ref{eq: FS K def}) and (\ref{eq: K<HS})
set a bound on the evolution speed:
\begin{equation}
\left|\frac{ds_{FS}}{dt}\right|=\sqrt{K}\le\left\Vert \mathcal{H}\right\Vert _{HS}.
\end{equation}
Upon writing the $HS$ norm explicitly in terms of matrix elements
\begin{equation}
\left\Vert \mathcal{H}\right\Vert _{HS}^{2}=\sum_{i,j}\left|\mathcal{H}_{ij}\right|^{2},\label{eq: HS comp}
\end{equation}
it becomes clear that the evolution speed is limited by the size of
the Hamiltonian elements and not just by the eigenvalue difference.
This becomes very important in the vicinity of NH degeneracies as
shown in Appendix 3.

\subsection{The spectral norm upper bound --- a tighter bound on the evolution
speed}

To get a tighter bound on the evolution speed we use the following
inequality:
\begin{equation}
K\le\la\Psi|\cH^{\dd}\cH|\Psi\ra=\sum_{k=1}^{N}\lb_{k}|\la\Psi|k\ra|^{2}\le\max(\lb_{k})\equiv||\cH||_{SP}^{2},\label{eq: K<SP}
\end{equation}
where $\lb_{k}\ge0$ and $|k\ra$ are the eigenvalues and eigenstates
of the matrix $\cH^{\dd}\cH$, \ $\cH^{\dd}\cH|k\ra=\lb_{k}|k\ra$.
$||\cH||_{SP}=\sqrt{\max(\lb_{k})}$ is known as the spectral norm
of $\cH$ (also known as Ky Fan 1-norm \cite{horn}).\textbf{ }To
understand the second inequality in (\ref{eq: K<SP}), notice that
the states $\{|k\ra\}$ constitute a complete orthonormal basis set
and $\braket{\Psi}{\Psi}=1$, so that the projection sum satisfies
$\sum_{k}\left|\left\langle \Psi|k\right\rangle \right|^{2}=1$. Thus,
$\sum_{k=1}^{N}\lb_{k}|\la\Psi|k\ra|^{2}$ is just a weighted average
of positive numbers $\lb_{k}$. Such a weighted average is always
smaller or equal to the largest element.

Obviously, the Hilbert-Schmidt norm $||\cH||_{HS}^{2}$ and the spectral
norm $||\cH||_{SP}^{2}$ can be represented in terms of the eigenvalues
$\lb_{k}$\,. This implies the following useful relation:
\begin{equation}
||\cH||_{SP}^{2}\le||\cH||_{HS}^{2}\le\mbox{rank}(H)||\cH||_{SP}^{2}\,.\label{fub-8}
\end{equation}
Therefore, the spectral norm bound on the evolution speed
\begin{equation}
\left|\frac{ds_{FS}}{dt}\right|=\sqrt{K}\le\left\Vert \mathcal{H}\right\Vert _{SP}\label{SP bound}
\end{equation}
is always tighter than the Hilbert-Schmidt norm bound. Simple and
useful lower and upper bounds on $\left\Vert \mathcal{H}\right\Vert _{SP}$
for $\cH(t)\in\CC^{N\times N}$ (but not on the evolution speed!)
are given by :
\begin{equation}
\max(\left|\cH_{i,j}\right|)\leq\left\Vert \cH\right\Vert _{SP}\le N\,\max(\left|\cH_{i,j}\right|).\label{eq: sp lower bound}
\end{equation}

Finally, we briefly comment on two extremal cases of traceless $2\times2$
matrix Hamiltonians.
\begin{itemize}
\item For a NH two-level Hamiltonian $\cH$ which is similar to a Jordan
block with zero-eigenvalue $\cH\sim J_{2}(0)=\left(\begin{array}{cc}
0 & 1\\
0 & 0
\end{array}\right)$ it holds $\mbox{rank}(\cH)=1$ so that
\begin{equation}
\cH\sim J_{2}(0)\qquad\Longrightarrow\qquad\left\Vert \mathcal{H}\right\Vert _{SP}=\left\Vert \mathcal{H}\right\Vert _{HS}.\label{eq: rank 1 case}
\end{equation}
This fact will be important later on in sec. \ref{sec: EP-DE}.
\item For a Hermitian traceless two-level Hamiltonian with energy separation
$\Delta E$, the spectral norm is $\left\Vert \mathcal{H}\right\Vert _{SP}=\left|\Delta E\right|/2$
and we obtain:
\begin{equation}
\left|\frac{ds_{FS}}{dt}\right|\leq\left|\Delta E\right|/2.
\end{equation}
As discussed in Appendix 1 the Bloch unit vector, $\hat{n}$, is related
to the Kinetic scalar via: $\left|\frac{d\hat{n}}{dt}\right|=2\sqrt{K}$.
Therefore the corresponding upper bound on the evolution speed over
the Bloch sphere in the \textit{Hermitian} case is:
\begin{equation}
\left|\frac{d\hat{n}}{dt}\right|=2\left|\frac{ds_{FS}}{dt}\right|\le|\Delta E|\label{fub-14}
\end{equation}
which is known as Fleming bound \cite{Fleming}. That is, for a two-level
Hermitian operator the spectral bound coincides with the known Hermitian
bound.
\end{itemize}
For explicit examples of the speed bound we refer the reader to Appendices
2 and 3. In Appendix 2, a Hamiltonian that describes certain optical
systems is analyzed. Appendix 3 studies a $\cP\cT$- symmetric Hamiltonian
that was introduced in \cite{cmb-brach-prl-2007}, in the context
of the $\cP\cT$-symmetric brachistochrone problem. In the next section
we introduce the notion of speed efficiency which quantifies how close
the actual motion in $\PP(\fH)$ is to the speed bound just derived.
Later we show how to construct a Hamiltonian which reaches the spectral
bound for a given motion in $\PP(\fH)$ at all times.

\section{Speed efficiency of quantum evolution\label{sec :Speed-Efficiency}}

In this section we introduce the notion of a maximally efficient evolution.
We wish to compare the actual speed of motion in the projective Hilbert
space $\PP(\fH)$ to the speed bound given by the spectral norm $\left\Vert \mathcal{H}\right\Vert _{SP}$
characterizing the available resources of the system. We use $\left\Vert \mathcal{H}\right\Vert _{SP}$
since it is tighter than the Hilbert-Schmidt norm $\left\Vert \mathcal{H}\right\Vert _{HS}$.
Let $\ket{\psi}\in\CC^{N}$ be a time-dependent state in an $N$-level
system that induces some predefined evolution $\pi(\ket{\psi})$ in
the corresponding projective Hilbert space $\PP(\fH)=\CC\PP^{N-1}\ni\pi(|\psi\ra)$.
We define the efficiency to be:

\begin{equation}
\eta(\mathcal{H},\ket{\psi})=\frac{\sqrt{K(\psi)}}{\left\Vert \mathcal{H}\right\Vert _{SP}}\le1\,.\label{eq: eta}
\end{equation}
It is important to realize that this efficiency is an instantaneous
(or local) property of $\mathcal{H}$ and its solution $\ket{\psi}$.
The shape of the curve in $\PP(\fH)$ alone has nothing to do with
efficiency. For example, a geodesic in $\PP(\fH)$ can have efficiency
smaller than one, and on the other hand, non-geodesic curves can have
100\% efficiency.

Loosely speaking, the value of $\eta$ quantifies to what extent the
Hamiltonian really uses all its resources to generate motion in $\PP(\fH)$.
That is why we call an $(\eta=1)-$evolution, a {}``maximally efficient
evolution''.\textbf{ }As an example of inefficient evolution, consider
a spin in a magnetic field which is not exactly perpendicular to the
spin direction. The part of the magnetic field which is parallel to
the spin is wasted as it does not contribute to the precession motion.
As we shall demonstrate in the next section, this inefficiency can
be fixed by making the Hamiltonian time-dependent (rotating the magnetic
field in time). In the NH case, reaching 100\% efficiency becomes
even more difficult. As explained at the end of section \ref{sec: Preliminaries},
for NH systems the condition $\frac{d}{dt}H=0$ does not guarantee
a constant evolution speed, i.e. $\left|\frac{ds_{FS}}{dt}\right|\neq\mbox{const}$.
On the other hand, the spectral norm is fixed if $\frac{d}{dt}H=0$.
Equation (\ref{eq: eta}), then, implies that $\eta$ varies with
time and, therefore, the evolution cannot be maximally efficient at
all times.

In the next subsection we show how to construct Hamiltonians that
are designed to generate maximally efficient evolution for a given
predefined motion in projective Hilbert space at all times. We will
demonstrate that such an $(\eta=1)-$evolution can be either Hermitian
or non-Hermitian.

\subsection{Maximally efficient evolution\label{sec: Maximal-Efficiency-Evolution}}

Our goal in this section is to find a Hamiltonian $\cH_{0}$ that
generates the same motion $\pi(\ket{\psi})$ in $\PP(\fH)$ as $\ket{\psi}$
but with 100\% efficiency. The solution $\ket m$ corresponding to
the maximally efficient Hamiltonian $\cH_{0}$ may differ from $\ket{\psi}$
only by a time dependent complex factor. In short, we look for $\ket m$
and $\mathcal{H}_{0}$ that satisfy:
\begin{eqnarray}
\pi(\ket m) & = & \pi(\ket{\psi})\label{req 1}\\
i\partial_{t}\ket m & = & \mathcal{H}_{0}\ket m\label{req 2}\\
\eta(t) & = & 1.\label{req 3}
\end{eqnarray}
The first requirement is that the states $\ket m$ and $\ket{\psi}$
have the same motion in $\PP(\fH)$. The second requirement states
that $\{\mathcal{H}_{0},\ket m\}$ satisfy the TDSE, whereas the third
requirement simply means that we are searching for maximal efficiency.
To satisfy the first requirement we set:
\begin{equation}
\ket m=c(t)\ket{\psi}\label{eq: |n>}
\end{equation}
where $c(t)$ is a complex differentiable function of time (see eq.
(\ref{eq: pi traj})). For reasons that will become clear shortly,
we fix $c(t)$ by choosing $\ket m$ to be normalized to unity and
to be parallel transported%
\footnote{If $\ket{\chi}$ is a normalized state, $\braket{\chi}{\chi}=1$,
then its parallel transported form is given by$\ket{\tilde{\chi}}=e^{-\intop_{0}^{t}\braket{\chi}{\partial_{t'}\chi}dt'}\ket{\chi}$.
\textbf{$\ket{\tilde{\chi}}$ }satisfies\textbf{$\braket{\tilde{\chi}}{\partial_{t'}\tilde{\chi}}=\braket{\tilde{\partial_{t'}\chi}}{\tilde{\chi}}=0$.
}This fixes the phase of the state up to a constant determined by
the choice of the lower limit of the time integral.%
}:
\begin{eqnarray}
\braket mm & = & 1,\label{eq: n norm}\\
\braket m{\partial_{t}|m}=(\partial_{t}\braket m{)|m} & = & 0.\label{eq: n phase}
\end{eqnarray}
The first condition determines $\left|c(t)\right|$, and the second
one yields the phase of $c(t)$ (up to a time-independent constant).
To find the Hamiltonian that drives $\ket m$ with maximal efficiency
we choose the following ansatz:
\begin{equation}
\mathcal{H}_{0}=i\ket{\partial_{t}m}\bra m-ig\ket m\bra{\partial_{t}m},\label{eq: H opt}
\end{equation}
where, in general, $g$ can be time-dependent. For $g=1$ the Hamiltonian
is Hermitian. Notice that $\ket{\partial_{t}m}\equiv\partial_{t}\ket m$
is not normalized and not parallel transported. In contrast to $\ket m$,
$\ket{\partial_{t}m}$ is also not a solution of the TDSE (with $\mathcal{H}_{0}$
as Hamiltonian). However, $\ket m$ and $\ket{\partial_{t}m}$ are
mutually orthogonal by virtue of the parallel transport we imposed.
By applying (\ref{eq: H opt}) to the state $\ket m$ we see that
the requirement (\ref{req 2}) is immediately satisfied. To fulfill
the remaining third requirement we note that in the basis $\{\ket m,\ket{\partial_{t}m}\}$
the Hamiltonian $\mathcal{H}_{0}$ has only off-diagonal elements
so that $\cH_{0}$ is traceless by construction. To calculate the
efficiency we\textbf{ }first calculate the spectral bound and then
the kinetic scalar\textbf{.} Evaluating $\mathcal{H}_{0}^{\dagger}\mathcal{H}_{0}$
we get
\begin{eqnarray}
\mathcal{H}_{0}^{\dagger}\mathcal{H}_{0} & = & \braket{\partial_{t}m}{\partial_{t}m}\ket m\bra m+\left|g\right|^{2}\ket{\partial_{t}m}\bra{\partial_{t}m}.\label{eq: H^dag H}
\end{eqnarray}
The eigenstates of $\mathcal{H}_{0}^{\dagger}\mathcal{H}_{0}$ are
$\ket m$ and $\ket{\partial_{t}m}$ and the corresponding eigenvalues
are $\braket{\partial_{t}m}{\partial_{t}m}$ and $\left|g\right|^{2}\braket{\partial_{t}m}{\partial_{t}m}$,
respectively.\textbf{ }The spectral norm is given by:
\begin{equation}
\left\Vert \mathcal{H}_{0}\right\Vert _{SP}=\sqrt{\braket{\partial_{t}m}{\partial_{t}m}}\max(1,\left|g\right|).
\end{equation}
Using the fact that $\braOket n{\mathcal{H}_{0}}n=0$ and Eq. (\ref{eq: FS K def})
and (\ref{eq: H^dag H}) we get that:
\begin{equation}
K(\mathcal{H}_{0},\ket m)=\braket{\partial_{t}m}{\partial_{t}m}.
\end{equation}
The efficiency, then, is given by:
\begin{equation}
\eta=\frac{\sqrt{K(\mathcal{H}_{0},\ket m)}}{\left\Vert \mathcal{H}_{0}\right\Vert _{SP}}=\frac{\sqrt{\braket{\partial_{t}m}{\partial_{t}m}}}{\sqrt{\braket{\partial_{t}m}{\partial_{t}m}}\max(1,\left|g\right|)}.
\end{equation}
Clearly, maximal efficiency $\eta=1$ (third requirement (\ref{req 3}))
is achieved provided that:
\begin{equation}
\left|g\right|\le1.\label{eq: g_cond}
\end{equation}

In summary, given any arbitrary state $\ket{\psi}$, equations (\ref{eq: |n>})-(\ref{eq: H opt})
together with (\ref{eq: g_cond}) show how to construct maximal efficiency
Hamiltonians that generate the same motion in projective Hilbert space
$\PP(\fH)$ as $\ket{\psi}$.

It is instructive to look on the instantaneous eigenvalues of $\mathcal{H}_{0}^{2}$.
From
\begin{eqnarray}
\mathcal{H}_{0}^{2} & = & g\ket{\partial_{t}m}\bra{\partial_{t}m}+g\braket{\partial_{t}m}{\partial_{t}m}\ket m\bra m\label{eq: H0_sq}
\end{eqnarray}
and $\mbox{tr}(\mathcal{H}_{0})=0$ it follows that the instantaneous
eigenvalues $E_{\pm}(t)$ of $\mathcal{H}_{0}$ are:
\begin{equation}
E_{\pm}(t)=\pm\sqrt{g}\sqrt{\braket{\partial_{t}m}{\partial_{t}m}}.
\end{equation}
In case of $g\in\RR$, these eigenvalues are real for $g>0$ and purely
imaginary for $g<0$, i.e. $E_{\pm}(t)\in\RR\cup i\RR$. This indicates
a hidden instantaneous pseudo-Hermiticity of $\cH_{0}$ --- in a certain
analogy to the considerations%
\footnote{We leave a corresponding detailed investigation to future research.%
} in Appendix 3. One of the key points of this work is that the evolution
can be maximally efficient regardless of whether the Hamiltonian is
Hermitian or not.

Finally, we note that the Hamiltonian $\cH_{0}$ in (\ref{eq: H opt})
shows some structural analogy to the brachistochrone Hamiltonians
for Hermitian systems (constructed in \cite{Hook}). In fact, $\cH_{0}$
extends the geodesic-trajectory paradigm of \cite{Dorje Herm,Hook,AA-prl-1990}
to maximally efficient evolution regimes over arbitrarily predefined
time-dependent trajectories in $\PP(\fH)$. Moreover, the constraint
$\Delta E=\mbox{const}$ is replaced by the constraint $\eta=1$.

\subsection{Inherent properties of the maximally efficient evolution}

Here we wish to highlight three points which are generic for maximally
efficient evolutions. The first point concerns the fact that $\ket m$
is normalized and parallel transported. In the construction of $\mathcal{H}_{0}$
we demanded $\braket{m(t)}{m(t)}=1$ and $\braket m{\partial_{t}m}=0$.
Now we wish to show that if these constraints are relaxed the spectral
norm will increase even though the motion in $\PP(\fH)$ remains unaltered.
According to (\ref{eq: eta}) the efficiency will drop below 100\%
by this modification. Assume we wish to change the amplitude and phase
of $\ket m$ by some complex factor $e^{-i\varphi(t)}$ where $\varphi(t)$
is some complex number. This is accomplished by adding $\mathcal{H}_{0}$
a diagonal term so that:
\begin{equation}
\mathcal{H}_{new}=\mathcal{H}_{0}+\ketbra mm\partial_{t}\varphi(t).
\end{equation}
To keep the trace zero, another diagonal term must be added as well,
in principle, but it is of no importance to the present discussion.
This transformation does not change the value of $K$. The spectral
norm squared is the largest expectation of $\mathcal{H}_{new}^{\dagger}\mathcal{H}_{new}$,
so:
\begin{equation}
\fl\left\Vert \mathcal{H}_{new}\right\Vert _{SP}^{2}\geq\braOket m{\mathcal{H}_{new}^{\dagger}\mathcal{H}_{new}}m=\left|\partial_{t}\varphi(t)\right|^{2}+\braket{\partial_{t}m}{\partial_{t}m}>\braket{\partial_{t}m}{\partial_{t}m}
\end{equation}
Since $K$ remained the same and spectral norm increased, we see that
the efficiency is now:
\begin{equation}
\eta_{new}=\frac{\sqrt{\braket{\partial_{t}m}{\partial_{t}m}}}{\left\Vert \mathcal{H}_{new}\right\Vert _{SP}}\le\frac{\sqrt{\braket{\partial_{t}m}{\partial_{t}m}}}{\sqrt{\left|\partial_{t}\varphi(t)\right|^{2}+\braket{\partial_{t}m}{\partial_{t}m}}}<1,
\end{equation}
 This decrease in efficiency with respect to $\mathcal{H}_{0}$ expresses
the simple fact that changes in phase and/or amplitude also require
spectral norm resources from the Hamiltonian. In order to direct all
resources to motion in $\PP(\fH)$, any phase and amplitude changes
should be avoided.

The second point concerns the role of $g$. While the $\mathcal{H}_{0}$
found earlier conserves the norm of $\ket m$, it doesn't do so for
other initial states (with the exception of the Hermitian case $g=1$).
Moreover, while $\ket m$ evolves exactly in the same way for all
values of $g$, the evolution of other states strongly depends on
the value of $g$. This is demonstrated in the example shown in Figure
1. The state of interest in this example was chosen to be: $\ket m=\cos(t)\ket{\uparrow}+\sin(t)\ket{\downarrow}$.
We considered two different maximally efficient Hamiltonians. The
first one is Hermitian (g=1) and the other is not (g=-0.8). The Hamiltonian
is constructed using the recipe in Sec. \ref{sec: Maximal-Efficiency-Evolution}.
If the initial state is $\ket{\psi(t=0)}=\ket{m(t=0}=\ket{\uparrow}$
we observe that, as expected, both Hamiltonians generate the same
evolution. Yet, if $\ket{\psi(t=0)}=\ket{\downarrow}$, the different
Hamiltonians generate different evolutions and the effect of $g$
becomes apparent.

\begin{figure}
\includegraphics[width=9cm]{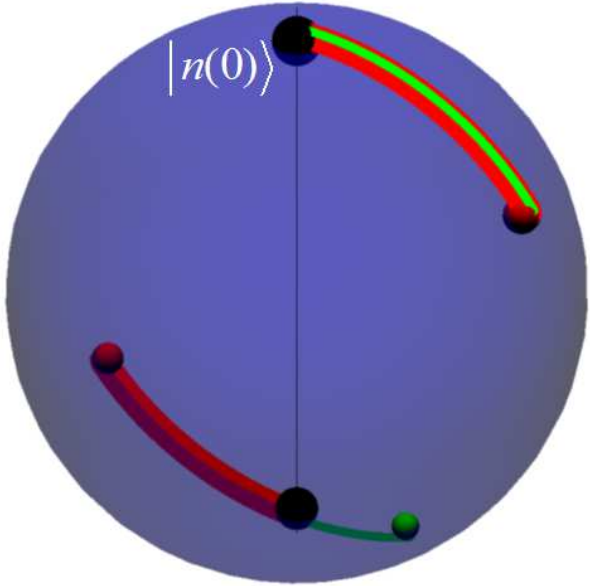}

\caption{(color online) The value of $g$ has no effect on the evolution when
applied to the initial state $\mathcal{H}_{0}$ was designed to propagate
efficiently (North pole of the Bloch sphere in this example). Yet
when applying $\mathcal{H}_{0}$ to a different initial state (South
pole) the $g$ may completely change the evolution. The large black
dots mark the starting points of the evolution. See text for details.}
\end{figure}

The third point that we note is that for periodic motion in $\PP(\fH)$
the state $\ket m$ accumulates only an Anandan-Aharonov phase \cite{AA phase},
since the dynamical phase $\braOket mHm$ is zero for maximally efficient
evolution. Once again, this is the result of wasting no resources
on phase accumulation.

\section{EP driven evolution\label{sec: EP-DE}}

A special case of great interest is $g=0$. Equation (\ref{eq: H0_sq})
shows that in this case $\mathcal{H}_{0}^{2}=0$. This can only happen
if $\mathcal{H}_{0}$ is similar to a rank-one Jordan block with zeros
on the diagonal, i.e. when $\cH_{0}\sim J_{2}(0)$. That is, $\mathcal{H}_{0}(g=0)$
describes an EP operator --- a NH degeneracy. The dynamics can be
fast even though at each instant the instantaneous eigenvalue difference
is zero. \textit{This degeneracy is time-dependent}. Moreover, the
orientation of the single geometric eigenvector of a Hamiltonian $\cH_{0}\sim J_{2}(0)$,
associates a preferred directionality to this degeneracy. The directionality
of the EP at each instant of time is chosen such that it induces the
desired dynamics. Thus it appears natural to name this evolution an
EP Driven Evolution (EP-DE). Another unique feature of this evolution
can be seen by evaluating the Hilbert-Schmidt norm. For a general
value of $g$ the $HS$ norm is:
\begin{equation}
\left\Vert \mathcal{H}_{0}\right\Vert _{HS}^{2}=(1+\left|g\right|^{2})\braket{\partial_{t}m}{\partial_{t}m}.
\end{equation}
Obviously, the $HS$ norm takes its minimal value for EP-DE ($g=0$),
i.e. from all the possible maximally efficient evolutions the EP-DE
has the minimal $HS$ norm. At this point the HS norm $\left\Vert \mathcal{H}_{0}\right\Vert _{HS}$
and the spectral norm $\left\Vert \mathcal{H}_{0}\right\Vert _{SP}$
coincide, a fact mentioned in (\ref{eq: rank 1 case}). Equation (\ref{eq: HS comp})
shows that the EP-DE provides the minimal value of $\sum_{i,j}\left|\mathcal{H}_{ij}\right|^{2}$
for a given trajectory in $\PP(\fH)$.

\section*{Conclusion}

The concept of speed efficiency was defined using spectral speed bounds
derived for NH or Hermitian systems. A recipe for the construction
of 100\% efficiency Hamiltonians for any evolution in the projective
Hilbert space was given. These Hamiltonians contain a free parameter,
$"g"$. 100\% efficiency is obtained for $\left|g\right|\le1$. The
Hermitian case corresponds to $g=1$. We conclude that it is possible
to have a Hamiltonian which is both 100\% efficient and Hermitian.
The $g=0$ case corresponds to a Hamiltonian which is not diagonalizable,
i.e. the evolution is driven solely by a time-dependent NH degeneracy
(exceptional point). This particular evolution minimizes the quantity
$\sum_{i,j}\left|\mathcal{H}_{ij}\right|^{2}$ with respect to all
other 100\% efficiency evolutions considered in this work.

\section*{Appendix 1 - The Bloch Sphere and the Fubini-Study metric}

The dynamics of NH $2\times2$ matrix systems in $\fH=\CC^{2}\ni|\psi\ra$
is conveniently analyzed as dynamics on the Bloch sphere. The latter
is spanned by the unit vectors
\begin{equation}
\hat{n}(t)=\frac{\left\langle \psi|\vec{\sigma}|\psi\right\rangle }{\left\langle \psi|\psi\right\rangle }=\left\langle \Psi|\vec{\sigma}|\Psi\right\rangle \in S^{2}\subset\RR^{3}.\label{eq: n def}
\end{equation}
Its close relationship to the projective Hilbert space $\CC\PP^{1}=\PP(\fH)=\CC_{*}^{2}/\CC_{*}\sim S^{2}$
can be seen by the explicit comparison of the Bloch sphere metric
with the Fubini-Study metric of $\CC\PP^{1}$. For a qubit $|\Psi\ra\in\CC^{2}$
parametrized as $|\Psi\ra=\left(\cos(\t/2),e^{i\phi}\sin(\t/2)\right)^{T}$,
$\t\in[0,\pi]$, \ $\phi\in[0,2\pi]$ it holds $\hat{n}=(\sin(\t)\cos(\phi),\sin(\t)\sin(\phi),\cos(\t))^{T}$
and the Bloch sphere metric reads
\begin{equation}
d\hat{n}^{2}=d\t^{2}+\sin^{2}(\t)d\phi^{2}.\label{fub-9}
\end{equation}
For the \textit{same state} $\ket{\Psi}$ the Fubini-Study metric
(\ref{eq: FS K def}) reduces to
\begin{equation}
ds_{FS}^{2}=\frac{1}{4}(d\t^{2}+\sin^{2}(\t)d\phi^{2})\label{fub-10}
\end{equation}
and therefore:
\begin{equation}
\left(\frac{d\hat{n}}{dt}\right)^{2}=4\left(\frac{ds_{FS}}{dt}\right)^{2}=4K(t).\label{eq: bloch speed}
\end{equation}

This is closely related to the fact that orthogonal states are antipodal
on the Bloch sphere having a geodesic distance $\pi$, whereas the
corresponding Fubini-Study distance as discussed in section 2 is $s_{FS}=\Theta=\pi/2$.

The main results of this article can be expressed using the Bloch
Sphere and the NH Bloch Equations (see for example \cite{MVB_pol})
formalism, but we found that the results are more neatly described
by the {}``ket-bra'' operator formalism and the Schrödinger equation.
Moreover, unlike the NH Bloch Equation formalism the {}``ket-bra''
formalism is applicable to a multilevel system without any alterations.

\section*{Appendix 2 - Speed bounds in optical systems\label{sec: optics examp}}

Let us examine the NH evolution in optical systems where the Hamiltonians
are explicitly known and a two-level description is either a good
approximation or even exact. Consider the Hamiltonian $\cH_{0}$ introduced
and studied in \cite{BU} in the context of {}``EP cycling'' \cite{Gilary,MVB_pol,WG,UMM}:
\begin{equation}
\cH(z)=\left(\begin{array}{cc}
0 & i\\
-iq(z) & 0
\end{array}\right),\label{eq: H Berry}
\end{equation}
where $z$, the propagation coordinate, plays the role of time. This
Hamiltonian can describe different physical systems. In \cite{WG},
it was used to describe the evolution of the transverse electric field
and its spatial derivative in a waveguide. In this system $q(z)$
is proportional to the change in the index of refraction with respect
to vacuum. In \cite{MVB_pol} $\cH(z)$ describes the evolution of
the two optical polarizations in crystals. $q(z)$ in this case is
related to the change in the transverse part of the reciprocal dielectric
tensor. The eigenvalue difference of $\cH(z)$ is $\Delta E=2\sqrt{q(z)}$,
and at $q=0$ a NH degeneracy (an EP) forms which is experimentally
well accessible in these systems.

From the structure of $\cH_{0}(z)$ in (\ref{eq: H Berry}), it is
obvious that for $q=0$ and $\Delta E(q=0)=0$ the evolution speed
does not vanish for all states. Using (\ref{eq: bloch speed}) and
the kinetic scalar definition (\ref{eq: FS K def}), one finds that
the non-vanishing angular velocity is $\mbox{\ensuremath{\left|\frac{d\theta}{dz}\right|}=}|\dot{\theta}|=2\left|q(z)\right|$
for the spin-up state $\ket{\uparrow}=(1,0)^{T}$, and $|\dot{\theta}|=2$
for the spin-down state $\ket{\downarrow}=(0,1)^{T}$. In particular,
at the degeneracy $q=0$ the spin-up state becomes an eigenstate ($\dot{\theta}=0$),
whereas the spin-down state still has a speed $\dot{\theta}=2$ regardless
of $\Delta E=0$. For $q\neq0$ the evolution speed of a general state
(not necessarily spin-up or spin-down) can be shown to be limited
by
\begin{equation}
\mbox{\ensuremath{\dot{\left|\theta\right|}}}\leq2\,\max(1,|q(z)|)\le2\,\sqrt{1+\left|q\right|^{2}}\,,\label{eq: H Br speed ineq}
\end{equation}
where the last inequality simply follows from $\max(\left|a\right|,\left|b\right|)\le\sqrt{\left|a\right|^{2}+\left|b\right|^{2}}.$
From the norms of the Hamiltonian (\ref{eq: H Berry}),$\left\Vert \mathcal{H}\right\Vert _{SP}=\max(\left|i\right|,\left|-iq(z)\right|)\,\,,\,\,\left\Vert \mathcal{H}\right\Vert _{HS}=\sqrt{1+\left|q\right|^{2}}$
, we see that
\begin{equation}
\mbox{\ensuremath{\dot{\left|\theta\right|}}}\le2\,\left\Vert \mathcal{H}\right\Vert _{SP}\leq2\,\left\Vert \mathcal{H}\right\Vert _{HS}.
\end{equation}
Hence, the maximal speed exactly fits within the spectral norm bound
of $\cH(z)$. Other optical systems whose evolution speeds near EPs
can easily be studied are discussed, e.g., in \cite{Morales}.

\section*{Appendix 3 - Spectral speed bounds for pseudo-Hermitian and $\cP\cT-$symmetric
two-level Hamiltonians\label{sub:Two-level-pseudo-Hermitian}}

We start from a general type NH traceless Hamiltonian, $\cH$, written
in terms of the Pauli matrices $\vec{\sg}$:
\begin{equation}
\cH(t)=[\vec{\alpha}(t)+i\vec{\beta}(t)]\cdot\vec{\sigma}\,,
\end{equation}
where $\vec{\alpha},\vec{\beta}\in\RR^{3}$ are some time-dependent
real vectors. The eigenvalues are:
\begin{equation}
E_{\pm}=\pm\sqrt{(\vec{\alpha}+i\vec{\beta})^{2}}=\pm\sqrt{\vec{\alpha}^{2}-\vec{\beta}^{2}+i2\vec{\alpha}\cdot\vec{\beta}},
\end{equation}
and become real or pairwise complex conjugate for
\begin{equation}
\vec{\alpha}\cdot\vec{\beta}=0\qquad\Longrightarrow\qquad E_{\pm}\in\RR\cup i\RR.\label{pseudo-cond}
\end{equation}
Operators and matrices with this specific spectral behavior are known
to be symmetric under an anti-unitary transformation \cite{BBM-jpa-2002},
to be pseudo-Hermitian \cite{ali-jmp-2002} and self-adjoint in a
Pontryagin space \cite{dijksma-langer-book-1996} (a finite-dimension
type version of a Krein space \cite{azizov-book-1989}).

For Hamiltonians $\cH$ with $\vec{\alpha}\cdot\vec{\beta}=0$ the
$SP$ norm and the eigenvalue difference reduce to
\begin{equation}
\left\Vert \cH\right\Vert _{SP}=|\vec{\alpha}|+|\vec{\beta}|\,,\qquad\Delta E=2\sqrt{\vec{\alpha}^{2}-\vec{\beta}^{2}}.\label{eq: sp of ph}
\end{equation}
Obviously it is possible to have an arbitrary large $\left\Vert \mathcal{H}\right\Vert _{SP}$
and a vanishing energy difference by choosing $|\vec{\alpha}|\to|\vec{\beta}|$.
This choice corresponds to an EP limit for which $\left|\Delta E\right|/\left\Vert \mathcal{H}\right\Vert _{SP}\to0$,
since the eigenvalues are very small near the degeneracy while $\left\Vert \mathcal{H}\right\Vert _{SP}$
remains roughly constant and finite.

A simple example for a NH Hamiltonian with a similar type of behavior
is the Hamiltonian $\cH$ used in studies of the $\cP\cT-$symmetric
quantum brachistochrone problem \cite{cmb-brach-prl-2007}
\begin{equation}
\fl\cH=\left(\begin{array}{cc}
ir\sin\chi & s\\
s & -ir\sin\chi
\end{array}\right)=s\sg_{x}+ir\sin\chi\,\sg_{z},\qquad r,s,\chi\in\RR.\label{eq: H bender}
\end{equation}
This complex symmetric Hamiltonian is $\cP\cT-$symmetric, $[\cP\cT,\cH]=0$
and $\cP-$pseudo-Hermitian, $\cP\cH=\cH^{\dd}\cP$, with the parity
operation given as $\cP=\sg_{x}$ and the time reversal, $\cT$, as
complex conjugation. In \cite{cmb-brach-prl-2007}, it was shown that
for certain parameter combinations $r,s,\chi$ such Hamiltonians with
fixed and purely real eigenvalues difference $\Delta E\in\RR$ can
evolve a given initial state $|\Psi_{I}\ra\in\CC^{2}$ into an orthogonal
final state $\ket{\Psi_{F}}\in\CC^{2}$, $\braket{\Psi_{I}}{\Psi_{F}}=0$,
in an arbitrarily short time interval. Due to the finite geodesic
distance $\pi$ between these antipodal states $\ket{\Psi_{I}}$ and
$\ket{\Psi_{F}}$ on the Bloch sphere, the corresponding evolution
speed should diverge in this limit. The concrete relations can be
easily obtained in terms of the simplifying reparametrization $r\sin\chi=s\sin\a$
which yields
\begin{eqnarray}
\fl\cH=s\left(\begin{array}{cc}
i\sin\a & 1\\
1 & -i\sin\a
\end{array}\right),\quad\Delta E=2s\cos\a,\quad\left\Vert \mathcal{H}\right\Vert _{SP}=|s|(1+|\sin\a|).\label{fub-14-1}
\end{eqnarray}
As demonstrated in \cite{GRS-ep-jpa-2007}, the ultra-fast evolution
regime predicted in \cite{cmb-brach-prl-2007} corresponds to an EP-limit
$\a\to\pm\pi/2$ so that for fixed $\Delta E=\mbox{const}$ it holds
$s=\Delta E/(2\cos\a)\to\infty$ and
\begin{equation}
\fl\cH\to s\left(\begin{array}{cc}
i & 1\\
1 & -i
\end{array}\right),\qquad\left\Vert \mathcal{H}\right\Vert _{SP}\approx2|s|\to\infty\,,\qquad\left|\Delta E\right|/\left\Vert \mathcal{H}\right\Vert _{SP}\approx\cos\a\to0.\label{fub-15}
\end{equation}
According to (\ref{eq: bloch speed}), this would indeed allow for
diverging evolution speeds on the Bloch sphere
\begin{equation}
\left|\frac{d\hat{n}}{dt}\right|=2\sqrt{K}\le2\left\Vert \mathcal{H}\right\Vert _{SP}\to\infty.\label{fub-16}
\end{equation}
From this diverging spectral norm one might be led to the conclusion
that actually such ultra-high evolution speeds and corresponding ultra-short
evolution times might be forbidden by the limited resources of the
system and the validity region of the model used. Both would set some
natural upper bounds (ultra-violet cut-offs) on the evolution speed.
This would be true if one were keeping within the present NH setups.
Nevertheless, the same ultra-high-speed evolution regimes can be induced
in subsystems of entangled Hermitian systems in larger Hilbert spaces
\cite{GS-brach-prl-2008}. Due to geometric contraction effects the
corresponding evolution speed of the associated (Naimark-dilated)
Hermitian system in the larger Hilbert space will remain finite, well-behaved
and much below any ultra-violet cutoffs.

For completeness we note that the present evolution speed considerations
are closely related to questions for possible lower bounds on evolution
times (quantum brachistochrone problems) and possible violations of
such bounds. Corresponding intensive theoretical studies for Hermitian
setups \cite{Dorje Herm,Hook,japan-brach-prl-2006} in the early 2000's
have been followed by investigations of various aspects of non-Hermitian
systems ($\cP\cT-$symmetric \cite{cmb-brach-prl-2007,cmb-lnp-2009,GRS-ep-jpa-2007,Ali-brach-prl-2007,Giri,GS-brach-pra-2008,GS-brach-prl-2008}
and quasi-Hermitian ones \cite{Ali-brach-pra-2009}, as well as other
of more general non-Hermitian types \cite{Assis}).

$\cP\cT-$symmetric setups \cite{cmb-prl-1998,cmb-rev-2007,ali-review-2010}
have been experimentally studied via special arrangements of gain-loss
components (active $\cP\cT-$symmetry) and components of different
loss (passive $\cP\cT-$symmetry) in optical waveguide systems \cite{christo-prl-09,christo-nature-10},
microwave billiards \cite{pt-micro-prl-2012}, electronic LRC-circuits
\cite{kottos-LRC-pra-11,kottos-UG-LRC-pra-12} and in mechanical systems
of coupled pendulums \cite{cmb-mech-exp-2012}.\\

\end{document}